\documentclass{article}

\usepackage{arxiv}
\usepackage[authoryear,longnamesfirst]{natbib}
\usepackage[utf8]{inputenc} 
\usepackage[T1]{fontenc}    
\usepackage{hyperref}       
\usepackage{url}            
\usepackage{booktabs}       
\usepackage{amsfonts}       
\usepackage{nicefrac}       
\usepackage{microtype}      
\usepackage{lipsum}
\usepackage{graphicx}
\graphicspath{ {./images/} }
\usepackage{siunitx}
\usepackage{float}      
\usepackage{placeins}   
\usepackage{enumitem}   
\usepackage{tcolorbox} 

\usepackage{amsmath,mathtools}
\usepackage[dvipsnames]{xcolor}
\definecolor{LightGray}{gray}{0.9} 

\usepackage{graphicx}
\usepackage{float}
\usepackage{subcaption}
\usepackage{booktabs}
\usepackage{threeparttable}
\usepackage{multirow}
\usepackage{array}
\usepackage{enumitem}
\usepackage{pgfplots}
\usepackage{pgfplotstable}
\usepackage{tikz}
\pgfplotsset{compat=1.18}
\usepackage{placeins}
\usepackage{tcolorbox}
\usepackage{listings}
\usepackage{siunitx}
\usepackage{xspace}
\usepackage{etoolbox}

\title{Directive, Metacognitive or a Blend of Both? A Comparison of AI-Generated Feedback
Types on Student Engagement, Confidence, and Outcomes}

\newcommand{\platformName}{RiPPLE\xspace}

\author{
Omar Alsaiari$^{1,2,*}$ \\
School of Electrical Engineering and Computer Science\\
The University of Queensland\\
St Lucia, Brisbane, QLD 4072, Australia\\
Department of Computer Science, College of Science and Arts\\
Najran University, Sharurah, Najran 68341, Saudi Arabia\\
\texttt{o.alsaiari@uq.edu.au} \\
\And
Nilufar Baghaei$^{1}$ \\
School of Electrical Engineering and Computer Science\\
The University of Queensland\\
St Lucia, Brisbane, QLD 4072, Australia\\
\texttt{n.baghaei@uq.edu.au} \\
\And
Jason M. Lodge$^{3}$ \\
School of Education\\
The University of Queensland\\
St Lucia, Brisbane, QLD 4072, Australia\\
\texttt{jason.lodge@uq.edu.au} \\
\And
Omid Noroozi$^{4}$ \\
Faculty of Education and Learning Sciences\\
Wageningen University and Research\\
Wageningen, The Netherlands\\
\texttt{omid.noroozi@wur.nl} \\
\And
Dragan Gašević$^{5}$ \\
Faculty of Information Technology\\
Monash University\\
Melbourne, Australia\\
\texttt{dragan.gasevic@monash.edu} \\
\And
Marie Boden$^{1}$ \\
School of Electrical Engineering and Computer Science\\
The University of Queensland\\
St Lucia, Brisbane, QLD 4072, Australia\\
\texttt{marieb@eecs.uq.edu.au} \\
\And
Hassan Khosravi$^{6}$ \\
Institute for Teaching and Learning Innovation\\
The University of Queensland\\
St Lucia, Brisbane, QLD 4072, Australia\\
\texttt{h.khosravi@uq.edu.au} \\
}

\begin{document}
\maketitle
\begin{abstract}
Feedback is one of the most powerful influences on student learning, with extensive research examining how best to implement it in educational settings. Increasingly, feedback is being generated by artificial intelligence (AI), offering scalable and adaptive responses. Two widely studied approaches are directive feedback, which gives explicit explanations and reduces cognitive load to speed up learning, and metacognitive feedback which prompts learners to reflect, track their progress, and develop self-regulated learning (SRL) skills. While both approaches have clear theoretical advantages, their comparative effects on engagement, confidence, and quality of work remain underexplored. This study presents a semester-long randomised controlled trial with 329 students in an introductory  design and programming course using an adaptive educational platform. Participants were assigned to receive directive, metacognitive, or hybrid AI-generated feedback that blended elements of both directive and metacognitive feedback. 
 Results showed that revision behaviour differed across feedback conditions, with Hybrid prompting the most revisions 
 compared to Directive 
 and Metacognitive.
 Confidence ratings were uniformly high, and resource quality outcomes were comparable across conditions. These findings highlight the promise of AI in delivering feedback that balances clarity with reflection. Hybrid approaches, in particular, show potential to combine actionable guidance for immediate improvement with opportunities for self-reflection and metacognitive growth.
\end{abstract}


\noindent\textbf{Keywords:}
Metacognitive Feedback; Directive Feedback; Hybrid Feedback; Generative AI; Self-Regulated Learning; Higher Education

\section{Introduction}

Feedback is consistently identified as one of the most powerful influences on student learning and achievement across educational contexts \citep{hattie2007power, wisniewski2020power}. Yet feedback is not a one-directional transfer of information; students actively interpret, evaluate, and decide how to act on it \citep{carless_development_2018, nicol2021power, doi:10.1080/02602938.2022.2059446}. This dynamic nature makes the design of feedback highly consequential, as it shapes not only immediate revisions, but also long-term engagement, confidence, and self-regulation of learning.  In practice, providing feedback that is both timely and personalised remains a major challenge, particularly in large classes or online environments where it is not feasible for instructors to deliver detailed comments to every student. This constraint has fuelled growing interest in \textit{AI-generated feedback} as a complement to teachers\textquotesingle{} work, offering scalable and consistent support while reducing instructor workload. Emerging evidence suggests that such feedback can positively influence learners\textquotesingle{} perceptions, actions, and outcomes \citep{ba2025unraveling,ALSAIARI2025105363}. Building on this promise, scholars have begun to articulate \textit{human-AI pedagogical frameworks} for feedback, which position AI as a collaborative partner whose contributions can be calibrated, coordinated, and contextualised to enhance the overall feedback process \citep{banihashem2025pedagogical}.

A central challenge in feedback design, whether provisioned by teachers or AI, lies in balancing clarity with reflection. One common approach is \textit{directive feedback}, which provides explicit corrections and prescriptive guidance. Such feedback offers immediate support by showing learners what is wrong and how to fix it, enabling even novices to quickly identify and remedy mistakes \citep{shute2008focus, wisniewski2020power}. Its value is particularly evident for beginners, as reducing the cognitive load of diagnosing errors allows working memory to focus on executing the task \citep{sweller2011cognitive,shute2008focus}. Directive feedback reliably accelerates early progress and reduces errors in constrained tasks where time is limited. Yet this efficiency carries risks: learners may comply superficially without understanding underlying concepts, become over-reliant on external guidance, and struggle to transfer knowledge to novel contexts \citep{nicol2006formative,shute2008focus,roll2018understanding}.

These limitations have prompted consideration of alternative approaches to feedback. One such approach is to ground feedback in Metacognitive and Self-Regulated Learning (SRL) theories, which together emphasise learners’ active control over their cognitive, motivational, and behavioural processes. Metacognitive theory focuses on developing both metacognitive knowledge-awareness of one\textquotesingle{}s strategies, strengths, and limitations and metacognitive regulation skills such as planning, monitoring, and evaluating one\textquotesingle{}s work \citep{flavell1979metacognition, schraw1995metacognitive, winne2017cognition}. Complementing this, SRL theory conceptualises learning as a cyclical process of forethought, performance, and self-reflection, through which learners set goals, monitor progress, and adjust strategies to achieve desired outcomes \citep{zimmerman2002becoming}. Feedback that incorporates reflective prompts seeks to strengthen learners\textquotesingle{} capacity to judge the quality of their work, calibrate their confidence, and take adaptive control of their learning \citep{efklides2006metacognition, panadero2017review}. By shifting the emphasis from simply informing students what to correct toward fostering awareness of how and why to revise, metacognitive feedback supports the development of transferable learning skills and long-term autonomy. At the same time, it presents challenges: reflective prompts can increase cognitive load and slow immediate revisions, particularly for novices under time pressure \citep{sweller2011cognitive, luo2023impact, de2025cognitive}, and their effectiveness depends heavily on students\textquotesingle{} feedback literacy-the ability to interpret, manage affect, and translate feedback into action \citep{carless_development_2018, boud2013rethinking}.

Although both directive and metacognitive feedback are well theorised, there is a lack of strong empirical evidence directly comparing their effects on learning. Most existing studies have been conducted in small cohorts or on narrow tasks, making it difficult to generalise findings across diverse learners and educational contexts \citep{zheng2024exploring, zhang2024effect, callender2016improving, sato2018metacognitive, aldino2025analytics}. Comparisons based on instructor-delivered feedback also face inherent challenges: outcomes may be confounded by differences in teachers’ expertise, communication style, and affective delivery, all of which can significantly shape how feedback is perceived and acted upon \citep{conrad2015examining, Brackett01122013}. These sources of variability limit the reliability of cross-context comparisons and obscure the specific contribution of feedback type itself. In this regard, \textit{AI-generated feedback} offers a powerful methodological advantage. Because AI systems can be systematically designed to deliver feedback in distinct type, they provide a means of ensuring consistency across large populations of students. This scalability and standardisation make it possible to rigorously evaluate the relative effects of different feedback approaches, disentangled from the idiosyncrasies of individual instructors.

This study addresses these gaps through a semester-long randomised controlled trial with 329 students in a university design and programming course. As part of the course, students completed an assessment task in the form of creating study resources (for example, drafting multiple-choice questions) and received AI-generated feedback on their work. Each student was randomly assigned to one of three feedback conditions: (1) in the directive condition, the AI provided straightforward corrections and direct suggestions for improvement; (2) in the metacognitive condition, the AI offered reflective prompts and questions designed to encourage students to think about their approach; and (3) recognising that directive and metacognitive feedback each have distinct benefits but also limitations, we introduced a hybrid condition that blended elements of both, consistent with calls in the feedback literature for designs that balance guidance and learner agency. Hybrid AI-generated feedback integrates directive clarity with reflective prompts, positioning them as complementary rather than competing strategies \citep{carless_development_2018,nicol2006formative,williams2024delivering,daumiller2025advancing}. Directive elements benefit novices and lower-achieving students, while reflective prompts are more effective for confident learners with stronger feedback literacy \citep{shute2008focus,butler2018toward}.

Guided by this design, our work addressed the following research questions.
\begin{itemize}[leftmargin=2em]
    \item \textbf{RQ1. Linguistic and structural features:} How do directive, metacognitive, and hybrid AI-generated feedback differ in linguistic and structural characteristics (e.g., length, complexity, imperatives, reflective prompts)? Establishing clear linguistic distinctions is essential to ensure that AI-generated feedback types are not only theoretically but also practically differentiable. Without this, any observed effects on learning may be confounded by overlap in textual form.
    
    \item \textbf{RQ2. Engagement with feedback:} How do AI-generated feedback types influence student engagement, as reflected in (a) interaction time, (b) feedback action on feedback , and (c) task-flow transitions? Engagement is a critical mediator between feedback and learning. Understanding how students act on different types of AI-generated feedback provides insight into whether it motivates meaningful revision or discourages sustained interaction.
    \item \textbf{RQ3. Confidence:} How do AI-generated feedback types affect students\textquotesingle{} confidence in the quality of their work after receiving feedback?  Confidence influences persistence, willingness to take risks, and future engagement with AI-generated feedback. Studying this dimension helps clarify whether feedback types support or undermine students’ self-belief and agency.
    \item \textbf{RQ4. Work quality:} To what extent do AI-generated feedback types affect the quality of students’ revised work? Ultimately, the value of feedback lies in whether it leads to improved outputs. Comparing effects on work quality provides evidence of which AI-generated feedback type best balances immediate improvement with longer-term skill development.
\end{itemize}

This study is the first field investigation in an authentic course setting to directly compare directive, metacognitive, and hybrid types of AI-generated feedback. By leveraging AI\textquotesingle{}s capacity for consistent delivery across large cohorts, we created a unique opportunity to evaluate these approaches side by side under real learning conditions. Drawing on student interaction logs, survey responses, and the quality of  work, we examine how AI-generated feedback types differ linguistically, how students engage with them, and how these interactions influence confidence and learning outcomes. In doing so, we move beyond theoretical assumptions and small-scale investigations to provide robust empirical evidence on the strengths and limitations of each feedback type, offering practical guidance for both the design of future AI-generated feedback systems and for educators seeking to integrate them into their teaching.

\section{Methods}
\subsection{Research Tool: The \platformName System}

\begin{figure}[h]
\centering
\includegraphics[width=1\linewidth]{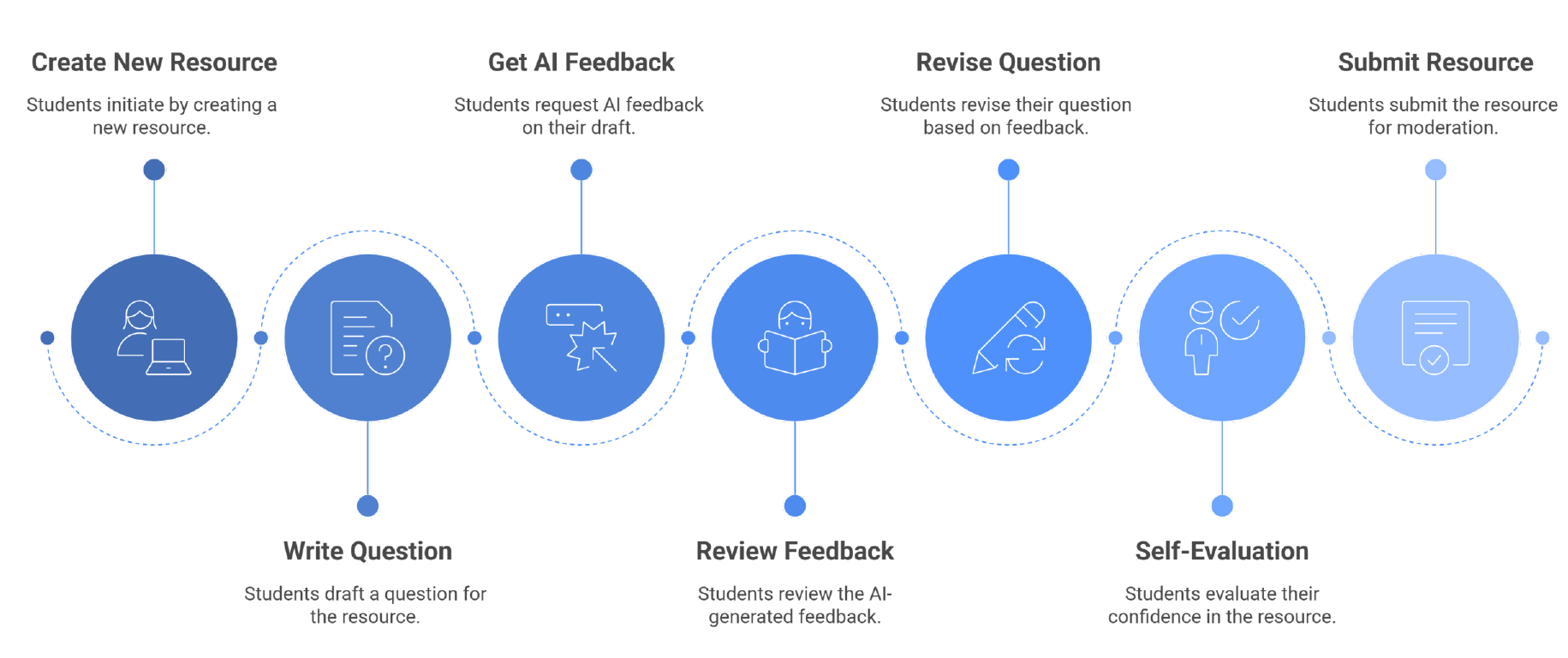}
\caption{Student resource creation process in \platformName}
\label{fig:placeholder}
\end{figure}

We employed \platformName, an adaptive educational platform designed to support student co-creation of learning resources \citep{khosravi2019ripple}. The system operates through a continuous cycle of creation, evaluation, and practice, with the creation process detailed in Figure~\ref{fig:placeholder}.
During the creation stage, students author educational resources including multiple choice questions, worked examples, and explanatory notes. This process requires deep engagement with course material as students identify key topics, formulate problems, and develop both correct solutions and plausible distractors.
Following creation, resources undergo peer evaluation where students assess the clarity, accuracy, and educational value of their peers\textquotesingle{}  work. While previous research examined various AI-generated feedback types within this evaluation phase \citep{ALSAIARI2025105363}, the current study specifically focuses on directive, metacognitive, and hybrid feedback conditions to investigate their distinct impacts.
The final stage involves students practicing with peer-approved resources. The system utilizes adaptive algorithms that model individual student knowledge and recommend resources tailored to their specific learning needs. Quality control is maintained through learning analytics and AI oversight, enabling instructors to effectively manage the process in large-scale educational settings.

This cyclical approach positions students as active educational content creators while simultaneously supporting their individual learning journeys. In this study, \platformName served as the experimental environment for implementing and evaluating directive, metacognitive, and hybrid AI-generated feedback interventions.

\begin{figure}
    \centering
    \includegraphics[width=0.9\linewidth]{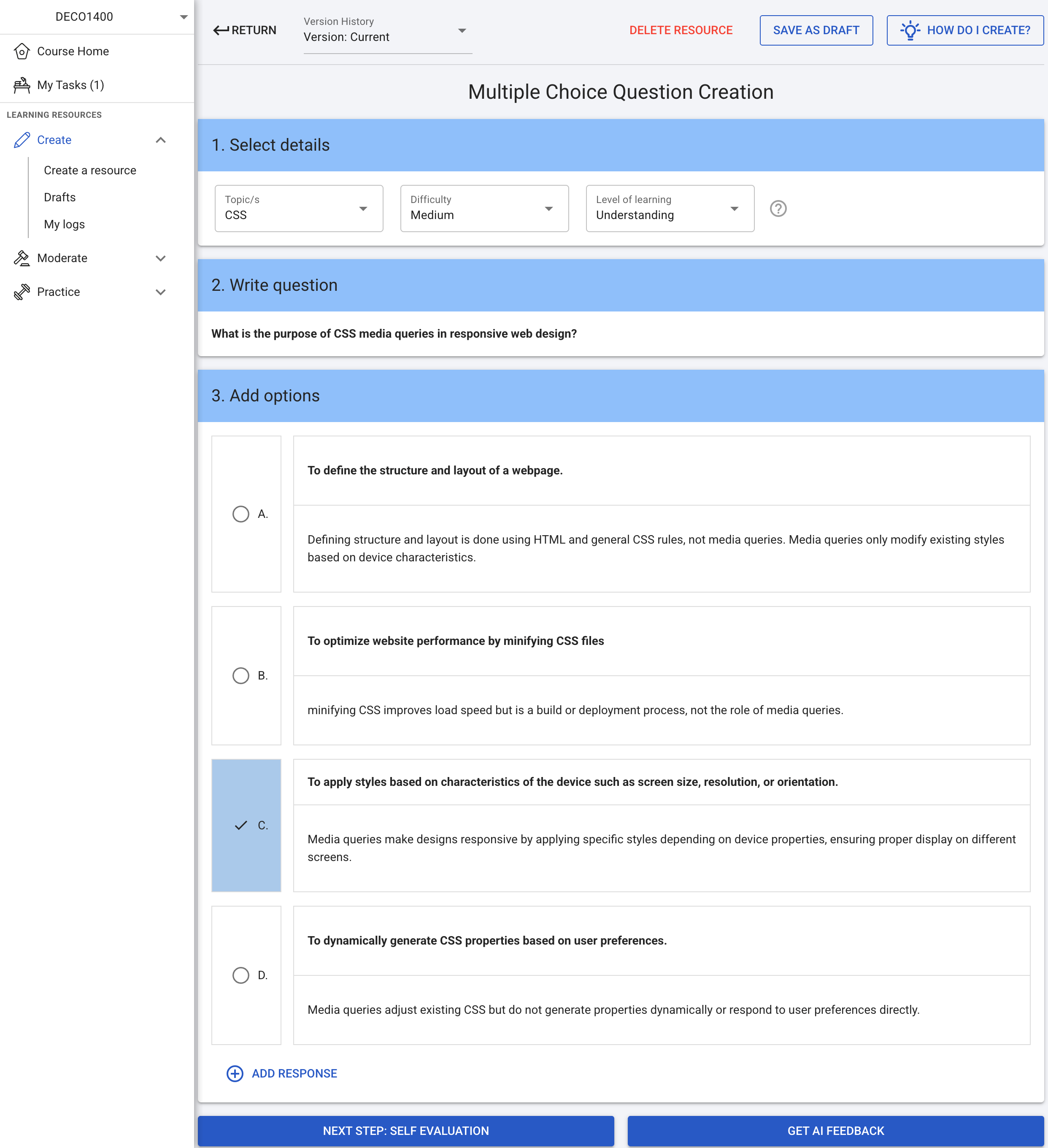}
    \caption{Creation stage on the RiPPLE platform showing the example MCQ used across all feedback conditions.}
    \label{fig:ripple_blind}
\end{figure}

\subsection{Conditions}
Three distinct types of AI-generated feedback were developed to investigate their effects on student responses: directive, metacognitive, and hybrid.

\textbf{Directive Feedback}: Participants received specific, actionable AI-generated feedback designed to directly improve their submissions. Feedback identified clear errors, offered concrete recommendations, and suggested precise revisions.

\textbf{Metacognitive Feedback}: Participants received AI-generated feedback structured to encourage reflective thinking and self-assessment rather than providing direct solutions. Students were prompted to consider various aspects of their submissions critically and independently, promoting deeper cognitive engagement.

\textbf{Hybrid Feedback}: This group received a balanced combination of directive and metacognitive AI-generated feedback. The prompts included explicit guidance for refinement as well as reflective considerations.\\

All AI-generated feedback types followed a fixed three-part structure: \emph{Summary}, \emph{Strengths}, and \emph{Suggestions for Improvement}. 
The \emph{Summary} and \emph{Strengths} sections were held constant in content and tone across all three AI-generated feedback types. 
Only the \emph{Suggestions for Improvement} section varied by condition.
The multiple-choice question used as the common example across all feedback conditions is shown in Figure~\ref{fig:ripple_blind}, and examples of the corresponding AI-generated feedback types are presented in Figure~\ref{fig:feedback_types_vertical}.

\subsection{Participants and Learning Context}

This study received ethical approval from the Human Research Ethics Committee at the university of Queensland [HE001453] and was conducted in a design and programming course. Of the 329 participants, 90.3\% were undergraduates, 6.1\% postgraduate students, and 3.6\% non-award students. International students comprised 67.5\% of the cohort, with 60\% studying English as a second language. Ages ranged from 18-33 years (M = 21).
Coursework tasks completed via the \platformName system contributed 10\% to final grades, through three four-week collaborative learning rounds. Each round required students to: (1) create one educational resource (multiple choice question, explanatory note, or flip card), (2) moderate five peer-authored resources, and (3) complete adaptive practice sessions using at least ten peer-created resources.
Students accessed \platformName online with instructions provided through lectures and announcements on the institutional learning management system (Blackboard).
\subsection{Study Design}
This randomised controlled study was conducted over one complete academic semester (12 weeks), providing sufficient time for meaningful learning interventions and behavioural pattern observation. Upon initial platform registration, participants were systematically randomly assigned to one of three experimental conditions: directive feedback, metacognitive feedback, or hybrid feedback groups, ensuring balanced distribution across conditions.
The study employed a between-subjects design to minimize cross-contamination effects and allow for clear comparison of intervention impacts. During the platform onboarding process, all students completed a  consent form, explicitly agreeing to the use of their learning data, platform interactions, and academic performance metrics for research purposes and system improvement initiatives.

\begin{figure}[htbp]
\centering

\begin{tcolorbox}[
  title=Directive Feedback,
  colback=white,
  colframe=black,
  colbacktitle=LightGray,   
  coltitle=black,          
  boxrule=0.4pt,
  arc=2mm,
  left=4pt,
  right=4pt,
  top=3pt,
  bottom=3pt
]
\begin{itemize}
  \item Refine the question to ensure it is meaningful without needing to read all the options first. A more direct phrasing such as \textit{"How do CSS media queries contribute to responsive web design?"} can make the intent clearer.
  \item Enhance the plausibility of the incorrect options by ensuring they are common misconceptions. For example, instead of \textit{"To optimize website performance by minifying CSS files,"} use \textit{"To dynamically generate CSS properties based on user preferences"} as a more realistic distractor.
  \item Improve consistency in the level of detail among options to avoid unintentional clues. Expand shorter distractors to match the complexity of the correct answer (e.g., \textit{"To apply styles based on user interaction and page loading speed"}).
  \item Ensure the explanations for incorrect options provide learning value. For instance, rather than stating \textit{"To define the structure and layout of a webpage"}, clarify why media queries go beyond static layouts.
\end{itemize}
\end{tcolorbox}

\begin{tcolorbox}[
  title=Metacognitive Feedback,
  colback=white,
  colframe=black,
  colbacktitle=LightGray,
  coltitle=black,
  boxrule=0.4pt,
  arc=2mm,
  left=4pt,
  right=4pt,
  top=3pt,
  bottom=3pt
]
\begin{itemize}
  \item Consider whether the question remains fully comprehensible without reading the options. Could slight rewording enhance clarity for those unfamiliar with media queries?
  \item Reflect on the phrasing of the distractors-do they all belong to the same conceptual category, or might one stand out as less related?
  \item Think about how the explanations for incorrect answers guide learners. Could an alternative approach further prompt students to analyze why their assumptions might be incorrect?
  \item Consider whether the explanation for the correct answer could be expanded to connect media queries with real-world examples of responsive web design.
\end{itemize}
\end{tcolorbox}

\begin{tcolorbox}[
  title=Hybrid Feedback,
  colback=white,
  colframe=black,
  colbacktitle=LightGray,
  coltitle=black,
  boxrule=0.4pt,
  arc=2mm,
  left=4pt,
  right=4pt,
  top=3pt,
  bottom=3pt
]
\begin{itemize}
  \item Refine the explanation for option D to provide a more detailed justification. Expand on how media queries work by referencing specific CSS rules (e.g., \texttt{@media} and common breakpoints), which would reinforce learning.
  \item Adjust the phrasing of option A to make the misconception more subtle. Instead of describing general CSS behaviour, consider referencing a different responsive technique such as flexbox or grid to better challenge understanding.
  \item Consider whether the question fully captures the depth of media queries\textquotesingle{}  functionality. How might adjusting the wording prompt students to explore other capabilities like device orientation or user preferences?
  \item Reflect on the balance of complexity among the options. Do all distractors require equal levels of critical thinking? If not, adjust their phrasing to distribute difficulty more evenly.
\end{itemize}
\end{tcolorbox}

\caption{Comparison of directive, metacognitive, and hybrid AI-generated feedback types}
\label{fig:feedback_types_vertical}
\end{figure}

\subsection{Data Analysis}

All analyses were conducted in R (version 4.4.0). We analyzed each research question with methods aligned to the outcome scale and the structure of the data. Descriptive statistics (means, medians, standard deviations) were computed for all variables prior to inferential analyses. An alpha level of .05 was used throughout. Outcome distributions were assessed with Shapiro-Wilk tests and visual inspection of Q-Q plots and histograms, which indicated non-normality and skewness. Where omnibus tests were significant, Tukey-adjusted pairwise contrasts based on estimated marginal means (EMMs) were conducted.

\subsubsection {RQ1. Linguistic and structural features }

All analyses were restricted to the \emph{Suggestions} segment of each AI-generated feedback text. 
This focus was chosen because the \emph{Summary} and \emph{Strengths} sections were held constant across all conditions, whereas the \emph{Suggestions} segment was the only part systematically varied by feedback type. 
We operationalised three features:

\begin{enumerate}[label=(\alph*)]
  \item \textbf{Word count}: total tokens based on word-boundary matching. 
  This provided a straightforward measure of feedback length, reflecting the overall quantity of information presented to students. 
  
  \item \textbf{Imperatives per 100 words}: count of \emph{sentences} classified as directive, normalised by word count. 
  Imperatives directly capture the prescriptive, action-oriented quality of directive feedback and allow comparison independent of text length. 
  
  \item \textbf{Reflective prompts per 100 words}: count of \emph{sentences} classified as reflective, normalised by word count. 
  Reflective prompts represent the core functional marker of metacognitive feedback, and normalisation ensures fair comparison across feedback types of differing length.  
\end{enumerate}

We removed formatting and markup, normalised whitespace, and segmented sentences using terminal punctuation (. ! ?). 
This ensured clean, consistent text units suitable for automated analysis and reduced noise from formatting artifacts. 
The detector allowed for leading bullets (\texttt{-}/\textbullet)
 at sentence start to accommodate list-based feedback structures common in AI output. 
Counts of imperatives and reflective prompts were then normalised per 100 words to control for variation in feedback length, ensuring that the resulting frequencies represented relative style rather than raw text volume. 
This step allowed valid comparisons across groups without confounding effects from differences in overall feedback verbosity.

To operationalise these measures, we next defined what qualified as a directive or reflective sentence.
A sentence is marked directive (imperative) if it matches either:
\begin{itemize}[leftmargin=1.2em]
  \item \emph{Clause-initial base-form verb} from a curated list: \textit{revise, fix, add, remove, replace, ensure, clarify, explain, provide, list, cite, check, format, organize, improve, update, expand, shorten, reword, correct, align, define, justify, label, write, mention, highlight, simplify, trim, streamline, rephrase, paraphrase, structure, standardize, balance, verify, validate, select, choose, exclude, include, maintain, keep, focus, enhance}.\\
  
  \item \emph{Modal/phrase-based directives}: ``(you) must/should/need to~…'', ``make sure to~…'', ``be sure to~…'', ``ensure that~…'', and prohibitives (``do not'', ``don't'', ``never'').
\end{itemize}
A sentence is marked reflective if it matches any of:
\begin{itemize}[leftmargin=1.2em]
  \item \emph{Strict ``consider'' forms}: ``consider how/why/whether/if/the extent to which/ways in which~…'' (bare ``consider'' does \emph{not} count).
  \item \emph{Non-question reflective phrases}: ``reflect on~…'', ``think about~…'', ``ask yourself~…'', ``you might~…'', ``it may help to~…'', ``revisit/rethink/reassess/re-evaluate~…''.
  \item \emph{Interrogative cues (require ``?'' at sentence end)}: ``how might~…'', ``what if~…'', ``to what extent~…'', ``does this~…'', ``are all~…'', ``might some~…'', ``is there a way~…''.
\end{itemize}
Having established these coding rules, we clarified how counts were applied. Counting is \emph{per category per sentence} (a sentence contributes at most 1 to each category). Categories were evaluated independently; a sentence could be both directive and reflective if it legitimately satisfied both patterns. We normalised counts by total words to obtain rates per 100 words.

Finally, to test whether AI-generated feedback type influenced linguistic features, we linked coded measures directly to feedback type (Directive, Metacognitive, Hybrid). Since AI-generated feedback texts were independent at the resource level, feedback condition served as the independent variable. A one-way ANOVA per feature provided the most direct test of group differences. We also fitted mixed-effects models with student-level clustering, which produced the same overall pattern of results; given the independence of feedback texts and fixed feedback type, we report ANOVA results for parsimony. When omnibus \(F\)-tests were significant, we conducted Tukey HSD for pairwise comparisons. We report \(\eta^2\) as the effect size (variance explained by feedback type). Assumptions were checked via residual diagnostics and Levene\textquotesingle{}s test for homogeneity of variance. 

\subsubsection{RQ2. Engagement with feedback}

Engagement was assessed through three complementary measures: (a) time-on-task as a proxy for cognitive investment, (b) action on feedback as a direct indicator of behavioural response, and (c) task-flow transitions to capture how feedback influenced subsequent actions.
  
\paragraph{Engagement time.} 
Engagement time was calculated as the interval between receipt of AI-generated feedback and subsequent submission. To limit extreme values, times were Winsorised at the 5th and 95th percentiles. A Gamma generalized linear mixed model (GLMM) with a log link was fitted, with feedback group as a fixed effect and student ID as a random intercept, as engagement times were non-normally distributed and positively skewed. This model also accounts for repeated measures within students.

\paragraph{Action on feedback.} 
Action on feedback was defined as a binary outcome, whether students revised their resource after receiving feedback but before the first self-assessment (\emph{Revised}=1, \emph{Not Revised}=0). This operationalization was selected to capture the presence or absence of a revision response to AI-generated feedback, serving as a clear and interpretable indicator of engagement. While we recognize that action on feedback may differ in scope or type (e.g.,, minor edits versus substantial restructuring), our primary focus was on whether feedback led to any revision, irrespective of magnitude. A binomial GLMM with a logit link was applied, using feedback group as a fixed effect and student ID as a random intercept, as the outcome was binary and data were clustered by student, requiring a model that both handles binary responses and accounts for repeated measures.

\paragraph{Event-flow transitions.} 
To complement  action on feedback measure, we applied First--Order Markov Models (FOMMs) as they provide a clear, interpretable way to model immediate transitions between task states and align with our focus on how feedback shaped subsequent actions. States reflected the platform workflow: \textsf{Question}, \textsf{QuestionDetails}, \textsf{Options}, \textsf{SelfAssessment}, and \textsf{Submission}, with all submissions routed through self-assessment. For each feedback group, a transition matrix was estimated; edges with probability $<.05$ were suppressed, node sizes indicated visit frequencies, and edge thicknesses represented transition probabilities.

\subsubsection*{RQ3. Confidence}

Confidence was measured through students’ self-ratings of resource quality on a five-point ordinal scale (1-5), ranging from Very Low to Very High. Ratings were collected immediately after revision or, when no revision occurred, prior to submission.A cumulative link mixed model (CLMM) with a logit link was fitted, with feedback group as a fixed effect and student ID as a random intercept,as the outcome was ordinal and not normally distributed. The CLMM is well-suited for ordered categorical data and accounts for repeated measures within students. Odds ratios from the CLMM provide interpretable effect sizes for contrasts between feedback types.

\subsubsection*{RQ4. Resource outcome quality}

Resource quality scores (0-5) reflected the platform’s evaluation pipeline, which combined peer ratings, instructor moderation, and algorithmic adjustments. Scores were rescaled to the interval $(0,1)$ because the distribution was negatively skewed toward higher ratings. A Beta GLMM  with a logit link was fitted, with feedback group as a fixed effect and student ID as a random intercept, as this model is appropriate for continuous outcomes bounded between 0 and 5 and accounts for repeated measures within students.

\section{Results}

\subsection{RQ1. Linguistic and Structural Features}

\begin{figure}[tbp]
    \centering
    \includegraphics[width=\linewidth]{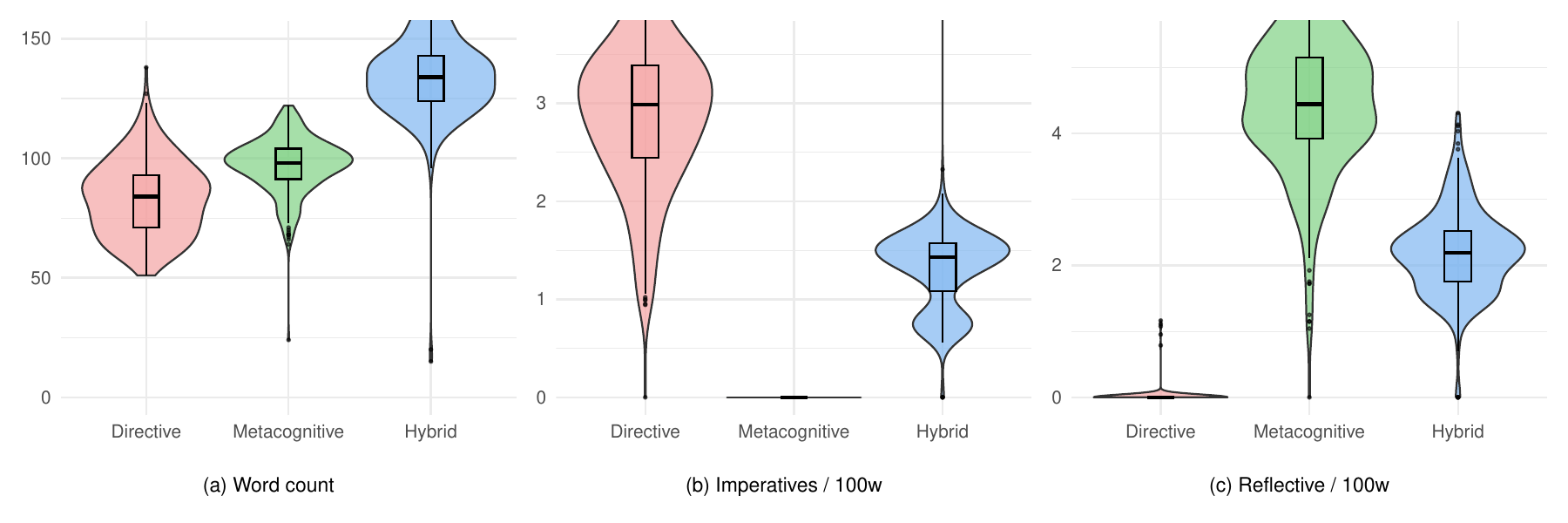}
    \caption{Distribution of linguistic features across feedback conditions.}
    \label{fig:rq1_violin}
\end{figure}

The analysis revealed substantial and systematic differences in linguistic and structural characteristics across the three feedback conditions, as presented in Table~\ref{tab:rq1_suggestions_structure_final_v2} and illustrated in Figure~\ref{fig:rq1_violin}.

\subsubsection*{Functional Markers}

Reflective prompts (per 100 words) varied strongly by condition. The Metacognitive condition exhibited the highest frequency (\textit{M} = 4.45, \textit{SD} = 1.07), the Hybrid condition showed intermediate levels (\textit{M} = 2.20, \textit{SD} = 0.65), and the Directive condition contained virtually none (\textit{M} = 0.01, \textit{SD} = 0.12). The omnibus test indicated a large effect, \textit{F}(2, 1024) = 3329.39, \textit{p} < .001, $\eta^2 = .867$. Tukey post hoc tests confirmed all pairwise comparisons as statistically significant (all adjusted \textit{p} < $10^{-12}$).

Imperative usage showed the opposite pattern. The Directive condition contained the highest frequency of imperatives per 100 words (\textit{M} = 2.91, \textit{SD} = 0.81), the Hybrid condition showed moderate use (\textit{M} = 1.34, \textit{SD} = 0.52), and the Metacognitive condition contained none (\textit{M} = 0.00, \textit{SD} = 0.00). The omnibus test was significant with a very large effect size, \textit{F}(2, 1024) = 2340.75, \textit{p} < .001, $\eta^2 = .821$, with all pairwise contrasts significant (all adjusted \textit{p} < .001).

\subsubsection*{Text Length Characteristics}

Word count also differed substantially across conditions. The Hybrid condition produced the longest text (\textit{M} = 134.14, \textit{SD} = 17.66, 95\% CI [132.23, 136.05]), the Metacognitive condition generated intermediate-length responses (\textit{M} = 97.10, \textit{SD} = 11.71), and the Directive condition yielded the shortest text (\textit{M} = 83.25, \textit{SD} = 15.63). The omnibus test again showed a large effect, \textit{F}(2, 1024) = 1019.56, \textit{p} < .001, $\eta^2 = .666$.

Tukey post hoc comparisons revealed significant differences between all conditions. The Hybrid condition exceeded the Directive condition by 50.89 words (95\% CI [48.17, 53.61]) and the Metacognitive condition by 37.04 words (95\% CI [34.28, 39.79]). The Metacognitive condition exceeded the Directive condition by 13.85 words (95\% CI [11.15, 16.56]). All contrasts were statistically significant (adjusted \textit{p} < .001).

\begin{table*}[ht]
\centering
\small
\setlength{\tabcolsep}{4pt}
\renewcommand{\arraystretch}{1.2}
\begin{threeparttable}
\caption{\textbf{RQ1: Linguistic and structural features of AI-generated feedback}}
\label{tab:rq1_suggestions_structure_final_v2}

\begin{tabular*}{\textwidth}{@{\extracolsep{\fill}} l l c c c c c c c}
\toprule
\textbf{Measure} & \textbf{Group} & $\mathbf{n}$ & \textbf{M} & \textbf{Mdn} & \textbf{SD} & $\mathbf{F(2, 1024)}$ & $\mathbf{p}$ & $\mathbf{\eta^2}$ \\
\midrule
\multirow{3}{*}{{Word count}} & Directive & 358 & 83.25 & 84.00 & 15.63 & \multirow{3}{*}{1019.56} & \multirow{3}{*}{$< .001$} & \multirow{3}{*}{.666} \\
 & Metacognitive & 338 & 97.10 & 98.00 & 11.71 & & & \\
 & Hybrid & 331 & 134.14 & 134.00 & 17.66 & & & \\
\midrule
\multirow{3}{*}{{Imperatives / 100w}} & Directive & 358 & 2.91 & 2.99 & 0.81 & \multirow{3}{*}{2340.75} & \multirow{3}{*}{$< .001$} & \multirow{3}{*}{.821} \\
 & Metacognitive & 338 & 0.00 & 0.00 & 0.00 & & & \\
 & Hybrid & 331 & 1.34 & 1.43 & 0.52 & & & \\
\midrule
\multirow{3}{*}{{Reflective / 100w}} & Directive & 358 & 0.01 & 0.00 & 0.12 & \multirow{3}{*}{3329.39} & \multirow{3}{*}{$< .001$} & \multirow{3}{*}{.867} \\
 & Metacognitive & 338 & 4.45 & 4.44 & 1.07 & & & \\
 & Hybrid & 331 & 2.20 & 2.19 & 0.65 & & & \\
\midrule[\heavyrulewidth]
\multicolumn{9}{l}{\textbf{Tukey Post-Hoc Comparisons (Mean Differences)}} \\
\toprule
\textbf{Measure} & \textbf{Comparison} & \multicolumn{2}{c}{\textbf{Mean Diff.}} & \multicolumn{2}{c}{\textbf{95\% CI}} & \textbf{Adj. $p$} & & \\
\midrule
\multirow{3}{*}{Word count} & Metacognitive -- Directive & \multicolumn{2}{c}{13.85} & \multicolumn{2}{c}{[11.15, 16.56]} & $< .001$ & & \\
 & Hybrid -- Directive & \multicolumn{2}{c}{50.89} & \multicolumn{2}{c}{[48.17, 53.61]} & $< .001$ & & \\
 & Hybrid -- Metacognitive & \multicolumn{2}{c}{37.04} & \multicolumn{2}{c}{[34.28, 39.79]} & $< .001$ & & \\
\midrule
\multirow{3}{*}{Imperatives / 100w} & Metacognitive -- Directive & \multicolumn{2}{c}{-2.91} & \multicolumn{2}{c}{[-3.01, -2.81]} & $< .001$ & & \\
 & Hybrid -- Directive & \multicolumn{2}{c}{-1.58} & \multicolumn{2}{c}{[-1.68, -1.48]} & $< .001$ & & \\
 & Hybrid -- Metacognitive & \multicolumn{2}{c}{1.34} & \multicolumn{2}{c}{[ 1.24, 1.44]} & $< .001$ & & \\
\midrule
\multirow{3}{*}{Reflective / 100w} & Metacognitive -- Directive & \multicolumn{2}{c}{4.43} & \multicolumn{2}{c}{[ 4.31, 4.56]} & $< .001$ & & \\
 & Hybrid -- Directive & \multicolumn{2}{c}{2.19} & \multicolumn{2}{c}{[ 2.06, 2.32]} & $< .001$ & & \\
 & Hybrid -- Metacognitive & \multicolumn{2}{c}{-2.24} & \multicolumn{2}{c}{[-2.37, -2.11]} & $< .001$ & & \\
\bottomrule
\end{tabular*}

\begin{tablenotes}
\small
\item Note. Descriptives are observed values. Omnibus tests are one-way ANOVAs; post hoc tests are Tukey HSD with adjusted $p$-values. Effect sizes ($\eta^2$) are for omnibus tests. Values rounded to two decimals.
\end{tablenotes}

\end{threeparttable}
\end{table*}

\subsection{RQ2: Engagement and Action on AI-generated feedback}

\subsubsection {Engagement Time}

A Gamma GLMM with a log link showed that the overall effect of feedback group was not statistically significant, $\chi^{2}(2) = 4.48$, $p = .107$.
EMMs indicated that students in the Directive group engaged for an average of 107.9 seconds (95\% CI [94.0, 124.0]), compared to 87.4 seconds (95\% CI [76.1, 100.0]) in the Metacognitive group and 97.8 seconds (95\% CI [85.3, 112.0]) in the Hybrid group.
Observed medians followed a similar pattern (Directive: 78.8 s; Metacognitive: 62.1 s; Hybrid: 72.9 s).
Pairwise Tukey comparisons revealed that Directive engagement times were approximately 23.5\% higher than Metacognitive ($p = .087$) and 10.4\% higher than Hybrid ($p = .579$), while Metacognitive was 10.6\% lower than Hybrid ($p = .496$).
However, none of these contrasts reached statistical significance.
Taken together, these findings suggest broadly comparable engagement times across the three feedback conditions.

\begin{table*}[ht]
\centering
\small
\setlength{\tabcolsep}{6pt}  
\renewcommand{\arraystretch}{1.15} 
\caption{\textbf{Engagement, confidence, and outcome}}
\label{tab:engagement_confidence_outcome}
\begin{threeparttable}
\centering
\begin{tabular*}{\textwidth}{@{\extracolsep{\fill}} l l c c c c c}
\toprule
\textbf{Measure} & \textbf{Group} & \textbf{n} & \textbf{Mdn (Observed)} & \textbf{Estimated Mean (EMM)} & \textbf{95\% CI} & \textbf{$p$-value} \\
\midrule
\multicolumn{7}{l}{\textit{Engagement Time (s)}} \\
 & Directive     & 332 & 78.8 & 107.9 & [94.0, 124.0] & \multirow{3}{*}{.107} \\
 & Metacognitive & 320 & 62.1 &  87.4 & [76.1, 100.0] &                       \\
 & Hybrid        & 315 & 72.9 &  97.8 & [85.3, 112.0] &                       \\
\midrule
\multicolumn{7}{l}{\textit{Confidence (1--5)}} \\
 & Directive     & 328 & 4.0 & 3.63 & [2.94, 4.32] & \multirow{3}{*}{.620} \\
 & Metacognitive & 322 & 4.0 & 3.25 & [2.48, 4.02] &                       \\
 & Hybrid        & 309 & 4.0 & 3.55 & [2.88, 4.23] &                       \\
\midrule
\multicolumn{7}{l}{\textit{Resource Outcome (1--5)}} \\
 & Directive     & 334 & 4.3 & 3.62 & [3.45, 3.78] & \multirow{3}{*}{.498} \\
 & Metacognitive & 326 & 4.3 & 3.72 & [3.55, 3.87] &                       \\
 & Hybrid        & 320 & 4.3 & 3.75 & [3.59, 3.91] &                       \\
\bottomrule
\end{tabular*}
\begin{flushleft}
\footnotesize
\textit{Note.} Mdn = observed median. EMM = estimated marginal mean from mixed-effects models (back-transformed to the original scale). 
$p$-values represent overall group effects from likelihood-ratio tests (Engagement: gamma GLMM; Confidence: CLMM; Outcome: beta GLMM).
\end{flushleft}
\end{threeparttable}
\end{table*}

\subsubsection {Action on feedback and Event-flow Transitions}

Action on feedback rates differed by feedback condition. Students in the Metacognitive condition acted least often (12.1\%), significantly less than those in both the Directive (21.1\%) and Hybrid (27.5\%) conditions, whereas rates did not differ significantly between the Directive and Hybrid conditions (see Table~\ref{tab:rq2_revision_adapted}).

\begin{figure}[tbp]
    \centering
    \begin{subfigure}{0.3\linewidth}
        \includegraphics[width=\linewidth]{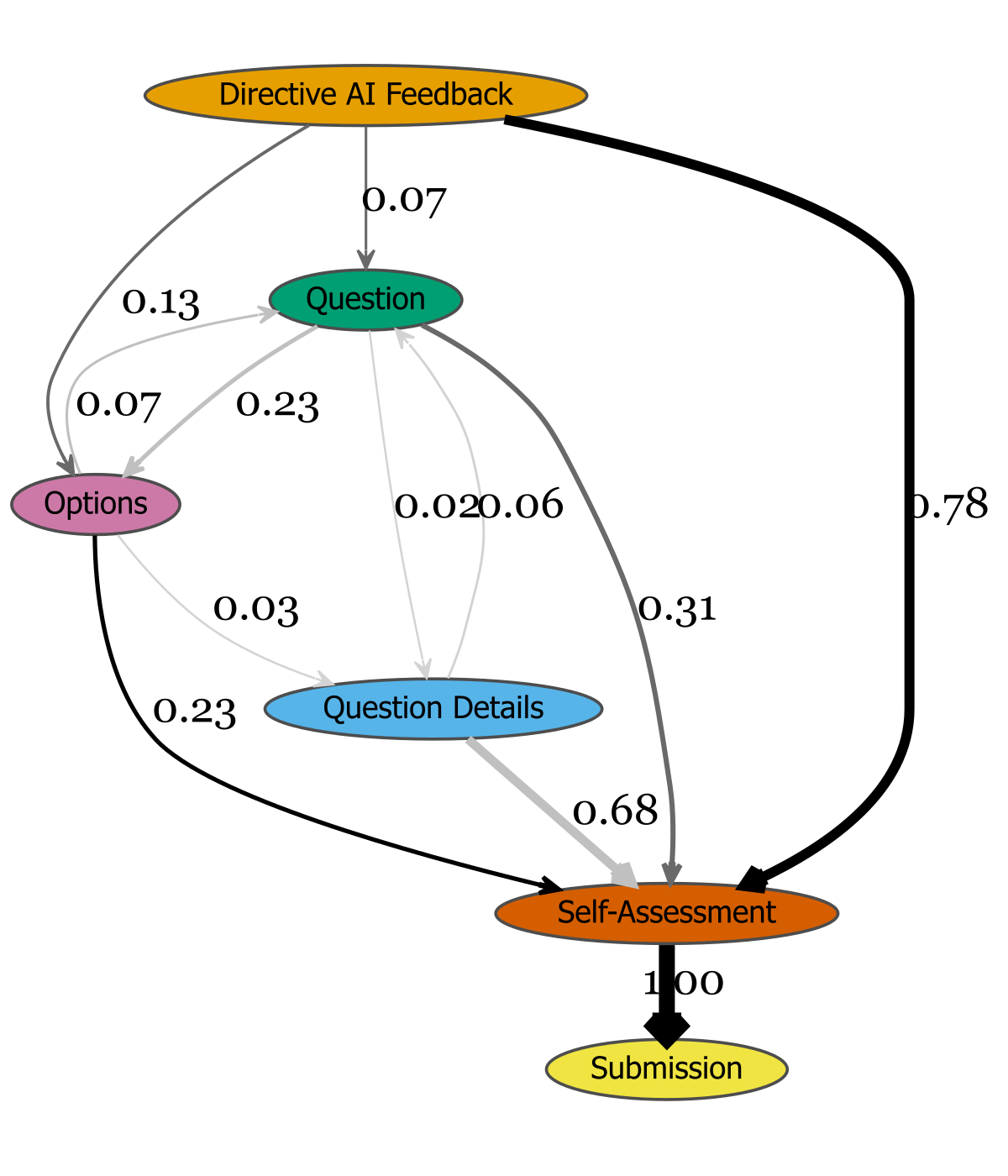}
        \caption{Directive Group}
    \end{subfigure}
    \hfill
    \begin{subfigure}{0.3\linewidth}
        \includegraphics[width=\linewidth]{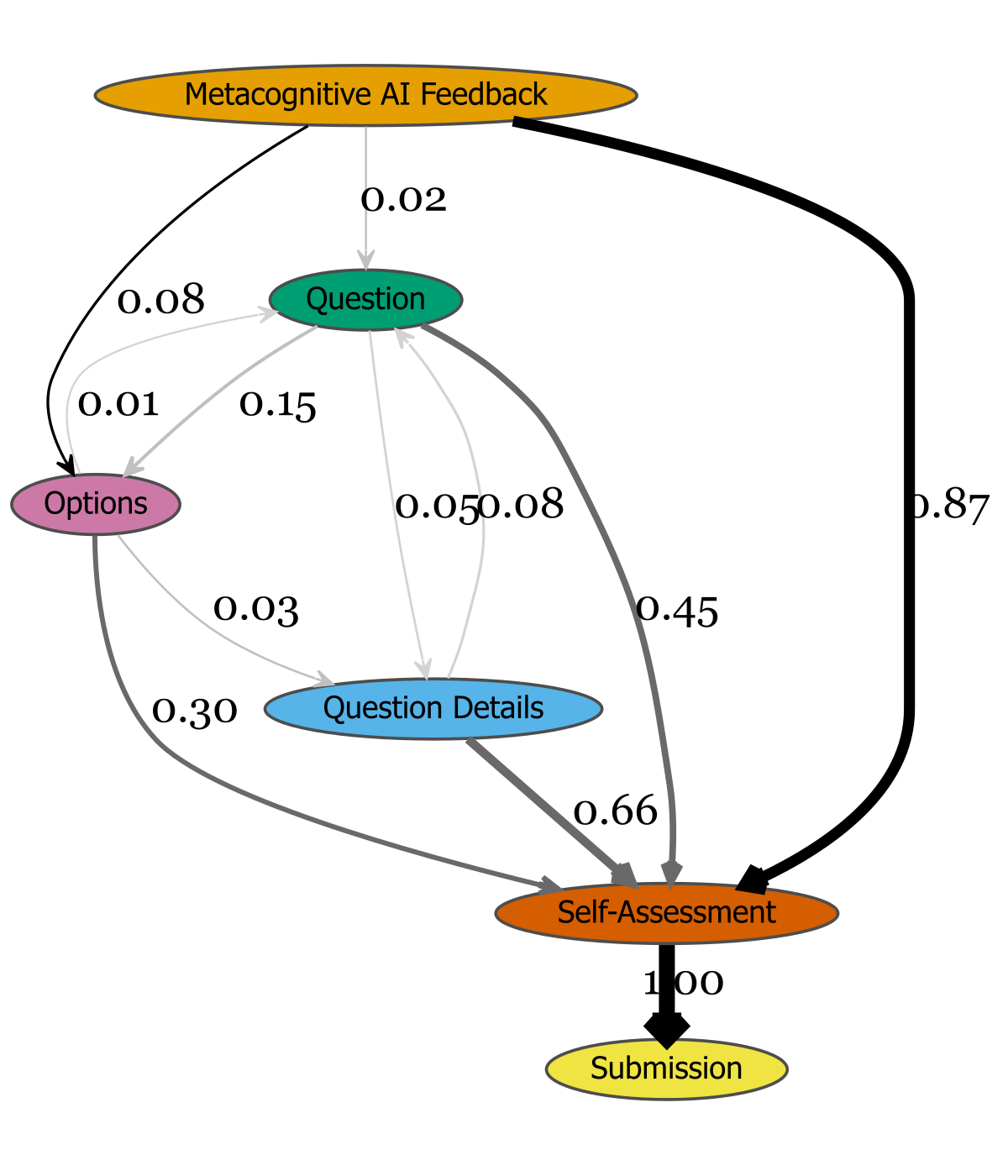}
        \caption{Metacognitive Group}
    \end{subfigure}
    \hfill
    \begin{subfigure}{0.3\linewidth}
        \includegraphics[width=\linewidth]{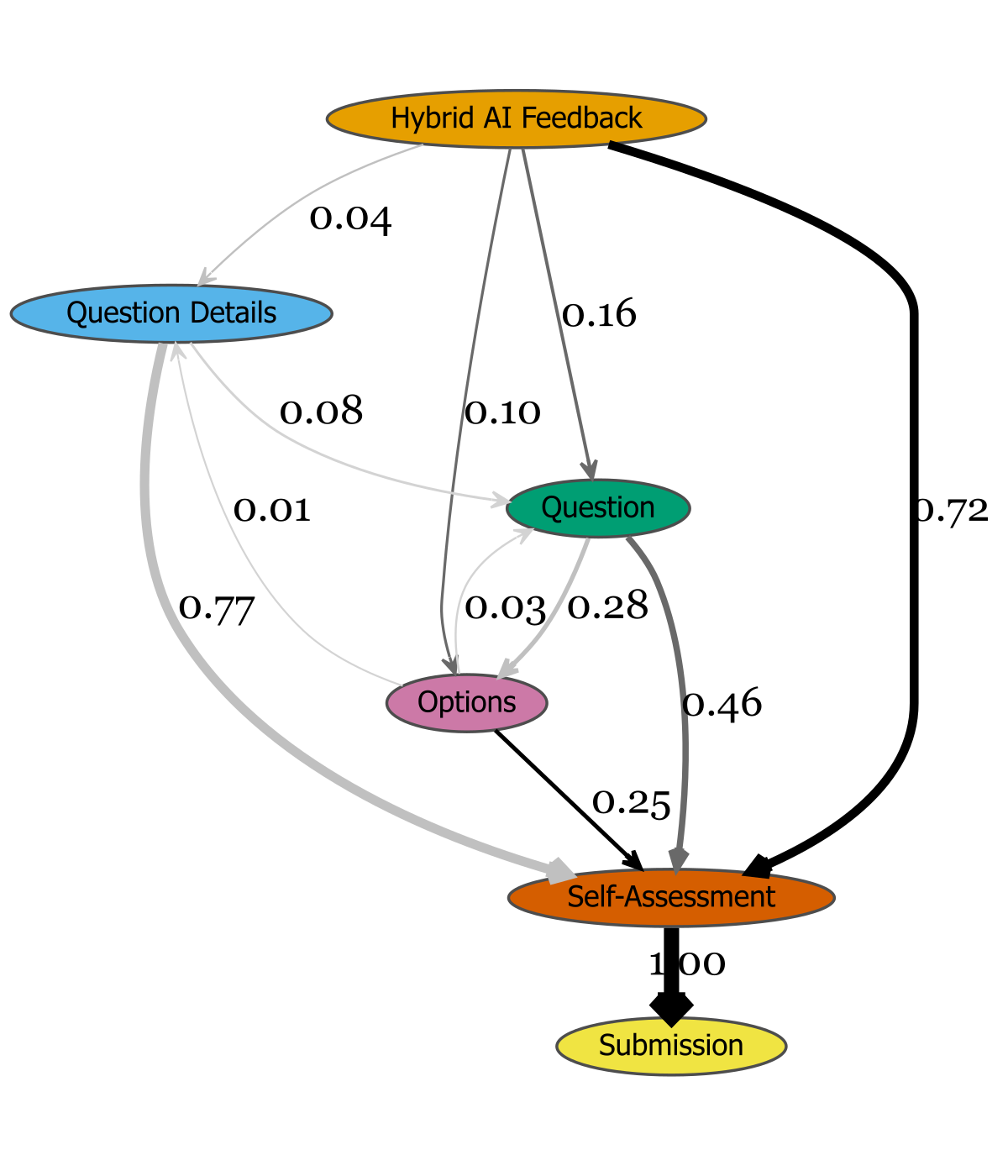}
        \caption{Hybrid Group}
    \end{subfigure}
    \caption{Task engagement flows across the three conditions (Directive, Metacognitive, Hybrid) using FOMM}
    \label{fig:FOMM_engagement}
\end{figure}

To contextualise these proportions with the task-flow traces, Fig.~\ref{fig:FOMM_engagement} panels~(a)--(c) display the event-flow graphs for each group from first AI-generated feedback onward, with all submissions routed through \emph{Self-Assessment}. Although all groups ultimately converged on the \emph{Self-Assessment} $\rightarrow$ \emph{Submission} pathway, the Hybrid group showed substantially more returns to editing before self-assessing. Specifically, 16\% of transitions from feedback led back to the \emph{Question} and 10\% to the \emph{Options}, compared with only 7\% and 13\% respectively in the Directive group, and 2\% and 8\% in the Metacognitive group. These denser editing loops in the Hybrid condition align with its higher observed action on feedback, whereas the Metacognitive condition shows relatively few returns to core editing states, consistent with its lower action on feedback.
A logistic mixed-effects model with random intercepts for students corroborated this ordering (Hybrid $>$ Directive $>$ Metacognitive) and indicated clear separation between the Metacognitive group and the other two. The overall effect of feedback condition was statistically significant, $\chi^2(2) = 8.74$, $p = .013$.

Tukey-adjusted pairwise contrasts indicated that students in the Directive group were nearly three times as likely to revise compared to those in the Metacognitive group (OR = 2.93, 95\% CI [1.25, 6.88], $p = .009$).The Hybrid group showed significantly greater action on feedback than the Metacognitive group (equivalently, Hybrid had about 4.3 times higher odds of revision; 95\% CI [1.82, 10.0], $p < .001$). No significant difference emerged between the Directive and Hybrid groups (OR = 0.69, 95\% CI [0.32, 1.48], $p = .484$). Taken together with the observed proportions in Fig.~\ref{fig:FOMM_engagement}, these results indicate  significantly lower action on feedback under Metacognitive feedback compared to both Directive and Hybrid feedback, with Hybrid showing the highest likelihood of revision.

\begin{table*}[ht]
\centering
\small
\setlength{\tabcolsep}{6pt}
\renewcommand{\arraystretch}{1.15}
\begin{threeparttable}
\caption{\textbf{RQ2: Action on feedback across feedback conditions (content-only)}}
\label{tab:rq2_revision_adapted}
\centering
\begin{tabular*}{\textwidth}{@{\extracolsep{\fill}} l c c c c c c c @{}} 
\toprule
\textbf{Group} & $\mathbf{n}$ & \textbf{\% Revised} & $\mathbf{M}$ & \textbf{Mdn} & \textbf{SD} & $\mathbf{\chi^2(2)}$ & $\mathbf{p}$ \\
\midrule
\multicolumn{8}{l}{\textit{Action on feedback }} \\
Directive & 331 & 21.1\% & 0.21 & 0 & 0.41 & \multirow{3}{*}{8.74} & \multirow{3}{*}{.013} \\
Metacognitive & 322 & 12.1\% & 0.12 & 0 & 0.33 & & \\
Hybrid & 306 & 27.5\% & 0.27 & 0 & 0.45 & & \\
\midrule
\textbf{Pairwise Comparison} & & & & \textbf{OR} & \textbf{95\% CI} & \textbf{Adj. $p$} & \\
\cline{5-7} 
Directive vs. Metacognitive & & & & 2.93 & [1.25, 6.88] & .009 & \\
Directive vs. Hybrid & & & & 0.69 & [0.32, 1.48] & .484 & \\
Metacognitive vs. Hybrid & & & & 0.23 & [0.10, 0.55] & $< .001$ & \\
\bottomrule
\end{tabular*}

\begin{tablenotes}
\small
\item Note. \% Not Revised is calculated as $100\% - \% \text{Revised}$. OR = Odds Ratio; CI = Confidence Interval; Adj. $p$ is the $p$-value adjusted for multiple comparisons. The omnibus $\chi^2$ test assesses the overall difference across the three conditions.
\end{tablenotes}
\end{threeparttable}
\end{table*}

\subsection{RQ3: Confidence}

A cumulative link mixed model (CLMM) with a logit link tested whether confidence ratings (1–5 scale) differed by feedback group. The overall effect of group was not statistically significant, $\chi^{2} = 0.96$, $p = .62$. The random intercept variance at the user level was 8.04 ($SD = 2.84$), indicating substantial between-student variability.

Estimated marginal means (back-transformed to the original 1–5 scale) suggested broadly similar confidence across groups: Directive ($M = 3.63$, 95\% CI [2.94, 4.32]), Metacognitive ($M = 3.25$, [2.48, 4.02]), and Hybrid ($M = 3.55$, [2.88, 4.23]). Observed medians were identical across groups ($Mdn = 4.0$).

Pairwise Tukey comparisons indicated that Directive confidence was about 10.5\% higher than Metacognitive ($p = .59$) and 2.2\% higher than Hybrid ($p = .96$). Hybrid was approximately 9.5\% higher than Metacognitive ($p = .71$). None of these contrasts reached statistical significance.

Overall, these findings indicate comparable confidence ratings across conditions, with a descriptive ordering of Directive $\approx$ Hybrid > Metacognitive.

\subsection{RQ4: Resource Outcomes}

A beta GLMM with a logit link showed that the overall effect of feedback group was not statistically significant, $\chi^{2} = 1.40$, $p = .49$. Estimated marginal means (back-transformed to the original 1-5 scale) indicated that students in the Directive group scored an average of 3.62 (95\% CI [3.45, 3.78]), compared to 3.72 (95\% CI [3.55, 3.87]) in the Metacognitive group and 3.75 (95\% CI [3.59, 3.91]) in the Hybrid group. Observed medians were the same for all AI-generated feedback groups.

Pairwise comparisons revealed that Metacognitive scores were approximately 3.0\% higher than Directive ($p = .40$), while Hybrid scores were about 4.1\% higher than Directive ($p = .25$). Hybrid was also 1.1\% higher than Metacognitive ($p = .81$). None of these contrasts reached statistical significance.

Overall, these findings indicate broadly comparable resource scores across the three feedback conditions, with a descriptive ordering of Hybrid > Metacognitive > Directive.

\begin{figure}[htbp]
  \centering
  \begin{subfigure}{0.33\textwidth}
    \centering
    \includegraphics[width=\linewidth]{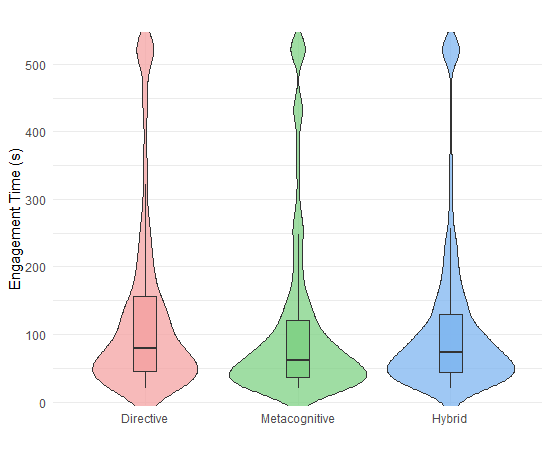}
    \caption{Engagement times}
    \label{fig:results:time}
  \end{subfigure}%
  \begin{subfigure}{0.33\textwidth}
    \centering
    \includegraphics[width=\linewidth]{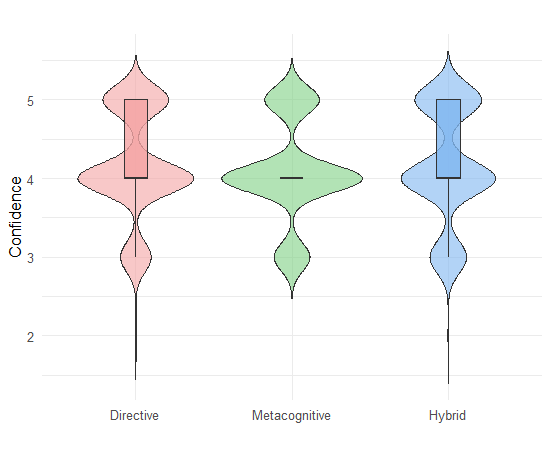}
    \caption{Confidence ratings}
    \label{fig:results:confidence}
  \end{subfigure}%
  \begin{subfigure}{0.33\textwidth}
    \centering
    \includegraphics[width=\linewidth]{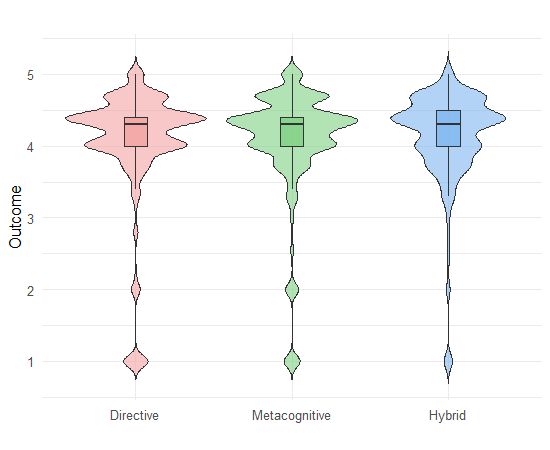}
    \caption{Resource outcomes}
    \label{fig:results:outcome}
  \end{subfigure}
  \caption{Results across feedback conditions: engagement, confidence, and outcomes.}
  \label{fig:results}
\end{figure}

\section{Discussion}

This study set out to examine how different AI-generated feedback designs, namely directive, metacognitive, and hybrid, shape learners’ engagement, revision behaviours, confidence, and performance outcomes. While prior work has demonstrated the potential of LLMs to generate pedagogically aligned feedback \citep{dai2024assessing, verma2024empirical}, the present findings provide a more fine-grained account of how specific linguistic and functional features of feedback map onto established theoretical frameworks such as Cognitive Load Theory, Metacognitive Theory, and Self-Regulated Learning. Taken together, the results highlight that feedback effectiveness cannot be captured solely by surface-level behavioural outcomes (e.g., time-on-task or revision quantity). Instead, in this study feedback effectiveness is conceptualized as a multidimensional construct reflected in cognitive (engagement time), strategic (action on feedback and task-flow transitions), and affective (confidence) dimensions. Below, we interpret the findings in relation to each research question and discuss their broader significance.

\subsection{Interpretation of the Results}

In response to \textbf{RQ1}, the results confirm that the feedback conditions were linguistically distinct in functionally meaningful ways. Directive feedback contained the highest frequency of imperatives, metacognitive feedback included the most reflective prompts, and hybrid feedback exhibited a balanced blend of both. These findings indicate that carefully crafted prompt design can produce functionally distinct AI-generated feedback types, reinforcing earlier claims that feedback should be intentionally engineered to align with pedagogical goals \citep{narciss2013designing, darvishi_impact_2024}. From a theoretical standpoint, the dominance of imperatives in directive feedback aligns with Cognitive Load Theory \citep{sweller2011cognitive}, which emphasizes reducing extraneous decision-making in novices through clear, actionable steps. This is echoed in recent findings by  Lee and Moore, who showed that generative AI feedback systems reduce instructional burden by offering concise, directive prompts that ease cognitive processing in learners  \citep{lee2024harnessing}. In contrast, the reflective phrasings characteristic of metacognitive and hybrid feedback support Metacognitive Theory, which emphasizes self-monitoring, evaluation, and regulation \citep{efklides2006metacognition, schraw1995metacognitive,lahza2025enhancing}. Critically, these findings confirm that the AI-generated feedback groups were meaningfully distinct in linguistic and functional form, ruling out the possibility that outcome differences stemmed from superficial lexical variation. Instead, they validate the intentional design of AI-generated feedback as a proxy for underlying instructional strategies.

In response to \textbf{RQ2a}, no statistically significant differences were found in engagement time across feedback conditions. At first glance, this may suggest that all feedback types triggered similar levels of time-on-task. However, this surface-level similarity conceals key differences in how time was used. In the metacognitive group, most students did not revise, yet spent as much time on the task as those in other groups who did revise. This indicates that their time was consumed by reading, interpreting, cognitively processing, and reflecting on the feedback. Such engagement is consistent with findings from \citep{liu2023second}, who noted that learners often spend considerable effort thinking through feedback, even when it does not lead to immediate revisions. This form of early-stage cognitive engagement aligns with Cognitive Load Theory and SRL theory, both of which emphasize that processing effort is not always reflected in behavioural output. The metacognitive group may have reached a cognitive “saturation point” after feedback analysis, leading to thoughtful inaction rather than revision, especially among novice learners \citep{ion2017enhancing}.

In response to \textbf{RQ2b\&c}, revision behaviour differed significantly across feedback types. The metacognitive group had the lowest revision action on feedback, whereas the hybrid group exhibited the highest. This supports the notion that metacognitive feedback, although cognitively demanding, often encourages learners to engage in strategic non-action, a well-documented SRL behaviour in which students choose to retain their original work after thoughtful evaluation \citep{zimmerman2002becoming}. However, cognitive overload may have hindered some students in the metacognitive group from translating reflection into action, particularly when confidence or clarity was lacking \citep{park2022l2}. Notably, the hybrid feedback group demonstrated high revision rates even when reflective prompts were included, indicating that combining directive and metacognitive elements could support both immediate revision and deeper cognitive processing. This aligns with work by \citep{torres2022feedback}, who found that dialogic feedback that balanced clear guidance with opportunities for reflection fostered co-regulation and metacognitive engagement.

In response to \textbf{RQ3}, confidence levels were consistently high across all feedback types, with no significant differences observed. This suggests that none of the feedback strategies undermined students’ belief in their work quality or in their own capabilities. Maintaining confidence is a critical concern in feedback design because negative emotional responses can inhibit motivation and learning \citep{kluger1996}. Instead, the stability of confidence across conditions points to the effectiveness of the feedback in supporting self-belief. One likely explanation is that all three feedback types contained positive elements that highlighted students’ strengths alongside areas for improvement. Prior research shows that balanced feedback that affirms competence while offering constructive guidance helps sustain confidence and fosters self-efficacy \citep{hall2007improving}. These findings also align with broader work linking self-efficacy and metacognitive development \citep{chen2019self, chung2021impact}.

In response to \textbf{RQ4}, despite differences in action on feedback across groups, there were no statistically significant differences in the quality of the resource outcome. A likely explanation is that students across conditions had already produced relatively high-quality drafts before receiving feedback, leaving limited room for observable improvement. This interpretation is consistent with ceiling effects in performance measurement, whereby additional feedback primarily serves to reinforce competence without producing large measurable gains \citep{bandura1997self}. Feedback in such contexts shifts from correction toward affirmation and consolidation, functions emphasized in both SRL theory \citep{zimmerman2002becoming} and feedback literacy frameworks \citep{carless_development_2018}. As  Li and Zhang showed, students often use feedback less to overhaul high-quality drafts and more to validate their revision goals and ensure alignment with academic expectations  \citep{li2021tracing}. Thus, the absence of differences in revision outcomes should not be interpreted as evidence of ineffectiveness; instead, feedback here served a role in refinement, validation, and self-regulatory development.

Another possible explanation concerns the source of feedback. Knowing it was AI-generated may have reduced trust compared to teacher feedback, making students more cautious in acting on it \citep{nazaretsky2024ai, henderson2025comparing}. The absence of outcome differences may reflect not ineffectiveness but feedback’s role in refinement, validation, and self-regulation, shaped by perceptions of credibility.

\subsection{Implications for Teaching and Learning}
From a pedagogical standpoint, hybrid AI feedback offers a practical means to address the enduring challenge of delivering timely, personalised feedback in large-enrolment and online courses. By combining directive guidance with reflective prompts, it accommodates varying levels of learner readiness without requiring instructors to have detailed insight into each student’s metacognitive development. This adaptability is particularly valuable in higher education contexts where feedback must remain scalable yet pedagogically meaningful.

For educators and teaching teams, effective integration of AI feedback requires more than adopting new technology. Professional development should cultivate a deep understanding of how AI-generated feedback is constructed, its pedagogical function, and its limitations. Without this foundation, instructors may struggle to align feedback with learning objectives or to intervene when students display counterproductive engagement behaviours.

Learners also benefit from explicit cultivation of feedback literacy. Because not all students possess the metacognitive readiness to translate reflective prompts into action, short scaffolded activities-such as modelling how to interpret and apply feedback-can bridge this gap. Such guided practice supports the development of self-regulation and evaluative judgement, encouraging learners to move from passive reception to active engagement with feedback.

Finally, embedding AI feedback within learning management systems enables data-informed pedagogy. Trace data on engagement duration, revision patterns, and confidence shifts can reveal when students read but fail to act on feedback, or when excessive revisions signal uncertainty. These analytics allow instructors to provide targeted, timely interventions that enhance both the effectiveness of feedback and the overall quality of learning outcomes.

\subsection{Limitations}

This study was conducted in a single course with relatively high baseline performance, which may limit generalizability and create ceiling effects. While the 12-week design allowed us to capture engagement across a full semester, we did not assess whether gains transferred beyond the course. Measures such as time-on-task and action on feedback served as behavioural proxies and may not reflect learners’ underlying cognitive processes, while self-reported confidence is subject to bias. Finally, we did not distinguish between surface edits and deeper integration of feedback, meaning some revisions may reflect compliance rather than genuine learning. Qualitative insights into students’ decision-making could have provided a richer account of how and why feedback was taken up.

\section{Conclusion}

This study systematically compared three distinct types of AI-generated feedback-directive, metacognitive, and hybrid within a higher education context to evaluate their respective impacts on student engagement, confidence, and submission  quality. The feedback types were deliberately engineered to exhibit clear linguistic and functional distinctions, ensuring robust differentiation between directive guidance, reflective prompting, and integrated approaches. This methodological rigor confirms the successful implementation of the experimental design and validates the integrity of the comparative analysis.
Examination of engagement patterns revealed nuanced differences in student response behaviors. While the engagement showed no significant variation across conditions, meaningful differences emerged in how students acted upon the feedback received. Students who received metacognitive feedback demonstrated the lowest revision rates, whereas those receiving directive or hybrid feedback were substantially more likely to engage in substantive revision activities. Notably, the hybrid condition yielded the highest overall revision rate, though the difference between directive and hybrid groups did not reach statistical significance. These findings suggest that feedback incorporating explicit, actionable guidance more effectively catalyzes student engagement with reflective prompts.
Confidence ratings were consistently high across all conditions, with no significant differences observed. Final resource quality scores also showed no significant variation between conditions. The hybrid approach, which integrates directive elements with metacognitive prompts, appears particularly promising for maximizing revision behavior while maintaining the reflective benefits of metacognitive scaffolding.
This study contributes empirical evidence demonstrating that thoughtfully designed AI-generated feedback can effectively support student engagement when strategically aligned with established pedagogical principles. These findings have significant implications for the integration of artificial intelligence tools in educational contexts, particularly regarding the design of automated feedback systems.

\subsection*{Declaration of Generative AI and AI-assisted technologies in the writing process}
During the preparation of this work, the main author used ChatGPT in order to improve the readability of the manuscript. After using this tool, the authors reviewed and edited the content as needed and take full responsibility for the content of the published article.

\FloatBarrier

\bibliographystyle{unsrt}
\bibliography{references}
\appendix

\end{document}